\documentclass[letter, 11pt]{article}
\usepackage{amsmath, amstext, graphicx, amsfonts, amssymb, mathrsfs, amsthm, amsopn, geometry, graphics, rotating, longtable, lscape, dsfont, multicol, pgfpages, xcolor, soul}
\usepackage[sectionbib]{natbib}

\definecolor{myblue}{RGB}{38, 28, 160}

\title{Who Flees Conflict?}

\author{Lidia Ceriani\thanks{lidia.ceriani@unibo.it}\\
\begin{minipage}[b]{5cm}\small \begin{center} University of Bologna\end{center} \end{minipage}\and  Paolo Verme\thanks{Corresponding author: paolo.verme@unibo.it. The authors wish to thank participants to the following meetings for their useful comments: seventh meeting of the Society for the Study of Income Inequality (ECINEQ) held in New York in July 2017; annual workshop of the Household in Conflict Network (HiCN) held in Brussels in November 2017; Georgetown Mortara International Development Seminar held in Washington DC in February 2018.  The authors also thank peer-reviewers at the World Bank, including Xavier Devictor, Toan Do,  Kevin McGee, David McKenzie and \c{C}ha\u{g}lar \"{O}zden. This work was originally part of the program ``Building the Evidence on Protracted Forced Displacement: A Multi-Stakeholder Partnership''. The program was funded by UK aid from the United Kingdom's Department for International Development (DFID), it was managed by the World Bank Group (WBG) and was established in partnership with the United Nations High Commissioner for Refugees (UNHCR). The scope of the program was to expand the global knowledge on forced displacement by funding quality research and disseminating results for the use of practitioners and policy makers. This work does not necessarily reflect the views of DFID, the WBG or UNHCR. }\\
\begin{minipage}[b]{5cm} \small \begin{center} University of Bologna \end{center} \end{minipage}}

\begin{document}
\thispagestyle{empty}
\maketitle
 \begin{abstract}
Despite the growing numbers of forcibly displaced persons worldwide, many people living under conflict \textit{choose} not to flee. Individuals face two lotteries - staying or leaving - characterized by two distributions of potential outcomes. This paper proposes to model the choice between these two lotteries using quantile maximization as opposed to expected utility theory. The paper posits that risk-averse individuals aim at minimizing losses by choosing the lottery with the best outcome at the lower end of the distribution, whereas risk-tolerant individuals aim at maximizing gains by choosing the lottery with the best outcome at the higher end of the distribution. Using a rich set of household and conflict panel data from Nigeria, the paper finds that risk-tolerant individuals have a significant preference for staying and risk-averse individuals have a significant preference for fleeing, in line with the predictions of the quantile maximization model. These findings are in contrast to findings on economic migrants, and call for separate policies toward economic and forced migrants.
\end{abstract}



\newpage

\setcounter{page}{1}

\section{Introduction}

The United Nations High Commissioner for Refugees (UNHCR) estimates that there were 117.3 million forcibly displaced people worldwide at the end of 2023.\footnote{http://www.unhcr.org/globaltrends} This estimate is the largest on record since the creation of the UNHCR in 1950 and is constituted by Internally Displaced Persons (IDPs, 68.3 m), refugees (43.3 m) and other people in need of international protection (5.7 m). The growth in these numbers is evidently associated with episodes of conflict and violence. Wars and civil conflicts in countries such as Syria, Afghanistan, Sudan, Ukraine, Democratic Republic of Congo, Nigeria, and Colombia among others have been responsible for major displacements of people internally and/or externally with no end in sight for many of these conflicts. 

Despite these numbers, most of the people who are affected by conflict do not migrate. Some people may have nowhere to go and others may not have the means or the opportunity to flee but many of those who live in conflict areas \textit{choose} not to move.\footnote{One study estimated that Conflict Affected Residents (CARs) accounted for about 87 percent of the total number of people affected by conflict (file:///C:/Users/wb201247/Downloads/PAC2013.pdf).} What are the factors that explain this choice remains one of the most puzzling and least researched questions in conflict and migration studies. 

In this paper, we posit that individuals living under conflict are Quantile Maximizers (QM) as opposed to Expected Utility Maximizers (EUM). Individuals face two lotteries - staying or leaving - corresponding to two distributions of outcomes that are observed and drive migration decisions. In an EUM framework, individuals focus on expected utilities (mean expected outcomes) and take decisions accordingly. In a QM framework, individuals self-select into groups that focus on different parts of the distributions of outcomes. In its most extreme form, we can imagine individuals partitioning into two groups. The first group focuses on the lowest outcomes of the two lotteries, and picks the lottery where this outcome is higher (\textit{maxmin} group). This is the group of individuals that exhibit the largest aversion to risk and prioritize minimizing losses. The second group focuses on the highest outcomes of the two lotteries and picks the lottery with the higher outcome (\textit{maxmax} group). This is the group of least risk-averse individuals (or risk-tolerant individuals) who seek to maximize gains. 

When compared to the EUM approach, the QM approach is more parsimonious, flexible and realistic. Individuals are not expected to be knowledgable of all possible prospected outcomes, since they focus on one particular segment of the distribution of outcomes. This is a more realistic assumption in a context where information is scarce and erratic. It is also not necessary to speculate on the shape of individual utility functions. If individual utilities are identical for all, the EUM model would not be able to predict who flees and who does not because all individuals would face the same expected utility and would take the same decision. If utilities vary across individuals, an EUM approach would require aggregating utilities across the population, which is complex. Instead, with a QM approach, individuals' risk preferences are revealed by the quantile of the distribution they focus on: at the extremes, \emph{maxmin} individuals are the most risk-averse and \emph{maxmax} individuals are the least risk-averse (risk-tolerant). This is simpler, closer to the reality of conflict, and also easier to test with empirical data.  

In our knowledge, this is the first paper that uses a QM approach to study risk preferences in a conflict scenario. The first paper that formalized a QM approach to decision making is \citet{Rostek_2010}\footnote{\citet{Chambers_2009} shows that quantiles are an essentially unique ordinal decision criterion which does not violate weak first-order stochastic dominance.}. The paper provided the theoretical characterization of quantile maximization in decision theory but did not offer any empirical application and we are not aware of empirical works that used this approach. The first and only paper we found that studies risk preferences as a possible driver of forced migration is \citet{Mironova_2017}. This is an unpublished experiment conducted in the Syrian Arab Republic and Turkey between 2013 and 2014. Using a variation of the \citeauthor{Eckel_2002} (2002 and 2008) risk game, the authors find combatants and non-combatants in rebel-held conflict zones to be more risk-tolerant than Syrians living in other parts of Syria or Syrians who fled abroad to Turkey. These results are consistent with our results although the paper does not build on a QM model.

Two papers looked at the role of conflict (or violence) in shaping risk preferences. Research across the social sciences has shown how traumas of different kinds (natural calamities, economic slumps, or personal traumas), or even the recollection of traumas, can modify risk preferences temporarily or permanently (\citealp{Malmendier_2011}; \citealp{Maarten_2012}; \citealp{Eckel_2009}; \citealp{Lerner_2001}). Similar findings are emerging in studies related to conflict. \citet{Maarten_2012} run a number of experiments among individuals who have been affected by violence in Burundi and find that individuals who experienced violence or lived in communities that experienced violence are significantly more risk-tolerant. \citet{Callen_2014} looked at violence and risk preferences in Afghanistan. They find that preferences for certain outcomes increase when individuals are exposed to violence or even recollect violent episodes (what they refer to as ``the certainty premium''). These studies are closely related to our work but they focused on understanding how conflict (or violence) affects risk preferences rather than understanding whether risk preferences contribute to select migrants and non-migrants. They are important for us because they point to a potentially confounding factor. Differences in risk preferences between migrants and non-migrants may be a reflection of the longer impact that conflict has on risk-aversion for the non-migrants rather than the outcome of a self-selection process, a hypothesis we will test in the paper.

The other body of literature that is related to our paper is the one on economic migration and risk preferences. This is quite vast and emerged in response to the seminal migration model proposed by \citet{Todaro_1969}. The original model focussed on rural-urban migration and identified expected income as the core driver of migration in line with EU theory. This hypothesis was later found to be inconsistent with some of the empirical evidence that emerged from developing economies and one of the early explanations for this failure was risk-aversion. \citet{Kuznets_1964} and \citet{Myrdal_1968} had already noticed that migrants are \textit{selectively} drawn from rural areas based on personal characteristics that are closely associated with risk-taking such as young age so that one should expect risk-tolerant individuals to be more likely to migrate. In a theoretical model devised as a critique to the Todaro model, \citet{Katz_1986} showed that a small chance of reaping high rewards is a sufficient condition to trigger migration irrespective of wage differentials, a finding that would make risk-taking behavior consistent with economic migration. Recent empirical work in developed and developing economies supports this view. Using data from the German Socio-Economic Panel (GSOEP), \citet{Jaeger_2010} find that Germans who had spells of migration between 2000 and 2006 were more likely to be risk-tolerant individuals as opposed to those who did not move. Using the 2009 round of the Rural Household Survey (RHS) of China, \citet{Akguc_2016} find that risk-tolerance positively affects the decision to migrate and this holds for households as well as for household heads and their spouses. It is important to note here that this literature focuses on self-selection of individuals into migrants and non-migrants. It does not address the question of whether exposure to poverty (instead of conflict) has an impact on risk preferences.\footnote{The theory and evidence on economic migration and risk preferences is more complex. For example, \citet{Bonin_2009} find that first generation immigrants to Germany have a higher risk-aversion than natives using the 2004 GSOEP, the same survey used by \citet{Jaeger_2010}. This is not necessarily in contrast with this latter paper because the samples compared are different but shows the importance of distinguishing between internal and external migration and between first and second generations of migrants. Some scholars have also provided arguments for economic migrants to be more risk averse. One hypothesis is that migration decisions are household based and that households strategically select individual members for migration in an effort to diversify and reduce risks (\citealp{Stark_1982}; \citealp{Katz_1986}). While plausible, there is not much empirical evidence supporting this hypothesis and such hypothesis is not necessarily in contrast with the fact that, within the household, risk-tolerant individuals would be more willing to migrate. Another hypothesis states that risks associated with the place of destination decreases over time as people settle and find work and that these risks may be below those of the place of origin, encouraging therefore risk averse individuals to migrate. Again, the empirical evidence for this hypothesis remains inconsistent as compared to the findings described in the text.} 

This paper uses household and conflict panel data from Nigeria for the period 2010-2016 to predict migration from conflict areas with a QM model. It finds the model to predict risk-averse individuals to leave conflict areas and risk-tolerant individuals to stay. Using a direct measure of risk preferences, it also finds these predictions to match observed risk preferences among leavers and stayers. If these results are confirmed, there is clearly a sharp difference between economic migrants and forced migrants in terms of risk preferences. Economic migrants are less risk-averse and migrate with the expectation of reaping large rewards and better living conditions. Forced migrants are more risk-averse and migrate in an effort to protect their lives and minimum living standards. As discussed in the conclusions, these findings have important policy implications for both the place of origin and the place of destination of migrants.
 
The paper is organized as follows. Section 2 summarizes traditional models of decision under uncertainty and outlines the model we propose; Section 3 describes the empirical strategy. Section 4 illustrates the data, Section 5 shows the results and Section 6 concludes summarizing results and discussing policy implications.

\section{Modeling Decisions under Uncertainty}
\subsection{Expected utility and recent developments}

Orthodox economic theory frames decision making under uncertainty using models of expected utility maximization.\footnote{The original idea dates back at least to Daniel Bernoulli (Versuch einer neuen Tizeorie der Wertbestirnrnung von Gliicksfdllen, Leipzig, 1896, translated by A. Pringsheim from ``Specimen theoriae novae de mensura sortis'' Commentarii acaderniae scienliarum imperialis Petropolitanae, Vol. V, for the years I730 and I731, published in 1738). It was rejected by classical economists such as Marshall on the ground that maximizing expected utility with a utility curve with diminishing returns implies that: \textit{``The gain in utility from winning a dollar will be less than the loss of utility from losing a dollar (...), individuals would always have to be paid to induce them to bear a risk. But this implication is clearly contrary to actual behavior''} (\citealp{Friedman_1948}).} The backbone of these models is the von Neumann-Morgenstern Expected Utility (EU) model (1944), where individuals' utilities over $n$ possible states of the world are described as

\begin{equation}\label{vNM}
U(x)=\sum_{1=1}^n \pi_i u(x_i)
\end{equation}

where $\pi_i$ is the probability that state of the world $i$ occurs, with $\sum_{i=1}^n \pi_i=1$, and $u(x_i)$ is the utility associated with outcome $x_i$, which we will consider some monetary value, such as levels of income or expenditure. Let $F_x$ denote the cumulative distribution function associated to the random variable (also referred to as prospect, or lottery) $x=\left[x_1, \pi_1; x_2, \pi_2, \dots; x_n, \pi_n\right]$, where $F_x=Pr(x\leq x_i)$, $i=1,2,\dots,n$.

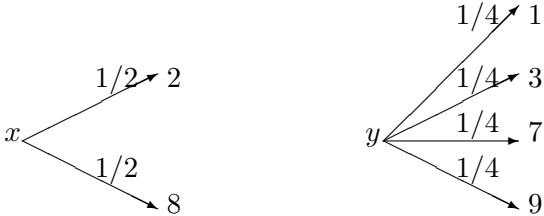
\begin{figure}[!htb]
\caption{Two lotteries $x$ and $y$, where $y$ is a mean preserving spread of $x$}\label{lottery_1}
\setlength{\unitlength}{1.2cm}
\begin{picture}(10,3.5)(-1,1)
\put(2.5, 2.5){\vector(2,1){1.5}}
\put(2.5, 2.5){\vector(2,-1){1.5}}

\put(2.3,2.5){$x$}
\put(3.3,3.1){$1/2$}
 \put(3.3,2.1){$1/2$}

\put(4.1,3.1){$2$}
\put(4.1,1.7){$8$}

\put(6.5, 2.5){\vector(1,1){1.5}}
\put(6.5, 2.5){\vector(2,1){1.5}}
\put(6.5, 2.5){\vector(1,0){1.5}}
\put(6.5, 2.5){\vector(2,-1){1.5}}

\put(6.3,2.5){$y$}
\put(7.3,3.8){$1/4$}
\put(7.3,3.1){$1/4$}
\put(7.3,2.6){$1/4$}
\put(7.3,2.1){$1/4$}

\put(8.1,3.8){$1$}
\put(8.1,3.1){$3$}
\put(8.1,2.5){$7$}
\put(8.1,1.7){$9$}
\end{picture}
\end{figure}

In standard models of these kinds, individuals are expected to be risk-averse so that the  function $u(.)$ is expected to be concave. A prospect $y$ is non preferred (riskier) to a prospect $x$ when $y$ is a \emph{mean preserving spread} of $x$ (see \citealp{Rothschild_1970}). Figure \ref{lottery_1} provides an example of a mean-preserving spread. Distributions $x$ and $y$ have the same expected outcome of $5$ but distribution $x$ has a smaller spread of outcomes than $y$. EU theory predicts that $U(x)>U(y)$, or that $x\succ y$ ($x$ is preferred to $y$). Risk-aversion induces individuals to choose the prospect with the lowest spread of outcomes (variance), given the same expected outcome. Note that for individuals to make a decision among two prospects, they must have complete information on the entire domain of the distribution of outcomes.

While the von Neumann-Morgenstern EU model remains the model of choice for most economists, several critiques have emerged over the past few decades. One early critique is the Ellsberg paradox (\citealp{Ellsberg_1961}) or the question of ambiguity aversion. When individuals are put in front of two alternatives, one with a certain low outcome and one with an uncertain high outcome, they would generally opt for the certain low outcome because of preferences for known odds to unknown odds, and this violates the postulates of expected utility theory. 

A second and more recent critique to the EU model is Kahneman and Tversky's [1979] \nocite{Kahneman_1979} prospect theory. The central theme of the critique is that people underweight outcomes that have very low probability of occurring and overweight outcomes that have a very high probability (this is called the \emph{certainty effect}). Considering equal weighting as in EU theory can lead to the Allais paradox (\citealp{Allais_1953}) where different choice frameworks can lead to opposite conclusions about the dominance of alternative choices. \citet{Kahneman_1979} showed that, for negative prospects, preferences are reversed as compared to positive prospects (this is called the \emph{reflection} effect). Therefore, ``\emph{ [...] certainty increases the aversiveness of losses as well as the desirability of gains}'' (\citealp{Kahneman_1979}, p. 269). The authors introduce the notion of decision weight to weigh the importance that people give to different probabilities so that the expected utility function becomes the following equation (\ref{KT}):

\begin{equation}\label{KT}
U(x) = \sum_{i=1}^n w(\pi_i) u(x_i).
\end{equation}

where the weighting function $w(\cdot): [0,1]\longrightarrow[0,1]$ infinitely underweighs infinitesimal probabilities and infinitely overweighs near-one probabilities.

Both the Ellsberg and Khanemann and Tversky critiques show that decisions under uncertainty cannot be driven simply by a mean and spread of a probabilistic distribution but that individuals attribute different importance to and have different preferences for different parts of the probability distribution. Behavioral and experimental economics have now provided countless examples supporting these views but it remains difficult to test the predictions of these models with non experimental and empirical data.  

Recent models of economic migration have also advanced our understanding of inter-temporal decisional processes emphasising the importance of past experiences and learning processes in decision making in a dynamic or life-cycle framework (\citealp{KenWal_2013}), a feature that is also shared by EU types of models that include subsistence thresholds. These models provide alternatives to classic EU models and allow for outcomes that could explain, for example, why individuals may choose not to migrate even when migrating would clearly improve their present welfare and utility.\footnote{In some cases, even material incentives to migrate have proved not to increase migration (see a recent review of this literature in \citet{BMY_2016}).} The general idea is that individuals have long-term rather than short-term goals and adapt to lower present consumption for ensuring long-term subsistence or other consumption goals. In the context of the Boko Haram conflict in North-East Nigeria and, more generally, in the context of forced migration, these models are less relevant as the decision to move is taken quickly and it is less likely that individuals can learn from past experiences. Indeed, one of the factors that explains Boko Haram's success is related to the fact that its incursions are audacious and unpredictable such as the kidnapping of 200 schoolgirls in 2014. Under such circumstances,  behavioral factors such as risk aversion become more important to decision making than rational planning based on past experiences.    

Our interest is to see how we can devise a model that draws on the intuitions provided by behavioral economics and the economics of migration but remains sufficently simple to predict migration decisions in the short-term and be tested with empirical data that are commonly available to researchers. The model presented in the next section is not necessarily a substitute of existing models but a viable alternative for empirical researchers working with real world data in conflict situations.

\subsection{A simplified Quantile Maximization Model}
We saw that in EU theory risk-aversion is a measure of how much individuals dislike the spread of prospect outcomes. We argue, instead, that a more relevant interpretation of risk-aversion for individuals facing conflict is their concern for \emph{downside risks}.

Consider the two prospects $x$ and $y$ reported in Figure \ref{lottery_1}, and assume that an individual is concerned not about the spread, but about falling below any given value with probability greater than $0.5$. With such (downside) risk of $0.5$, this individual obtains $2$ in lottery $x$ and $3$ in lottery $y$. Therefore, this individual will choose lottery $y$. Note that this different interpretation of risk-aversion (aversion to downside risk as opposed to spread), allows a more parsimonious assumption on the set of information available to individuals. In particular, individuals are not expected to have information about $F_x$ and $F_y$ over their entire domain, but only up to the smallest realization of the prospects $x$ and $y$ such that (in this specific example) $F_x\geq0.5$ and $F_y\geq0.5$. Also, no hypothesis on the shape of individual utility functions is needed. The only assumption is that people prefer more to less. 

If individuals are concerned with downside risk and not with mean preserving spreads, EU Maximization is not helpful in modeling and predicting individual choices. An alternative framework which models the attitude of individuals towards downside risk is \emph{Quantile Maximization} (QM). In QM, individuals concentrate only on selected segments (quantiles) of the distribution of prospective outcomes, according to their risk attitude. On one extreme, the most cautious individual will choose the lottery with the highest return in the worst case scenario (\emph{maxmin} decision maker); on the other extreme, the least cautious will pick the lottery with the highest return in the best case scenario (\emph{maxmax} decision maker). More in general, individuals may be somewhere in between the two extreme cases, being $\tau$-quantile decision makers.  

\citet{Rostek_2010} provides the theoretical characterization of quantile maximization in decision theory. Let $F_x$ be the cumulative probability distribution of the prospect $x$. Then, the $\tau$-th \emph{quantile} of $x$ is a (generalized) inverse of the cumulative distribution at $\tau$, defined as the smallest value $x_i$ such that the probability that the prospect $x$ will be less than $x_i$ is not smaller than $\tau$:

\begin{equation}
Q^\tau(F_x) = \inf\left\{ x_i | F_x \geq\tau \right\},
\end{equation}

while for $\tau = 0$, the quantile is defined as
\begin{equation}
Q^\tau(F_x) = \sup\left\{ x_i | F_x\leq 0\right\}
\end{equation}

A decision maker is said to be a $\tau-$\emph{quantile maximizer} if there exists a unique $\tau \in [0,1]$, such that for all $x,y \in \mathcal{F}$, 
\begin{equation}
x \succ y \iff Q^\tau(F_x) > Q^\tau(F_y).
\end{equation}

Similarly to the EU maximizer, where individuals evaluate lotteries on the basis of a single statistics, namely the mean, the $\tau$-maximizer assesses the value of each lottery by the $\tau$-th quantile realization. As mentioned above, at the extreme, quantile maximization includes the \emph{maxmin} maximizer who picks the lottery with the highest payoff among the lowest outcomes:

\begin{equation}
x\succ y \iff \min_{x_i|(\pi_i>0)}u(x) > \min_{y_i|(\pi_i>0)}u(y)
\end{equation}

and the \emph{maxmax} maximizer selects the lottery with the highest payoff among the highest outcomes:

\begin{equation}
x\succ y \iff \max_{x_i|(\pi_i>0)}u(x) > \max_{y_i|(\pi_i>0)}u(y)
\end{equation}

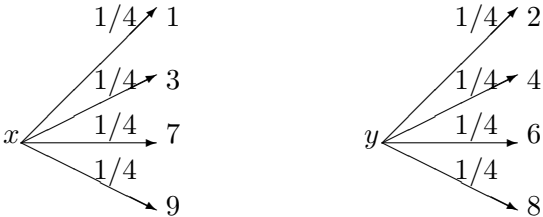
\begin{figure}[!htb]
\caption{Probabilities and outcomes associated to two lotteries $x$ and $y$}\label{lottery}
\setlength{\unitlength}{1.2cm}
\begin{picture}(10,3.5)(0,1)
\put(2.5, 2.5){\vector(1,1){1.5}}
\put(2.5, 2.5){\vector(2,1){1.5}}
\put(2.5, 2.5){\vector(1,0){1.5}}
\put(2.5, 2.5){\vector(2,-1){1.5}}

\put(2.3,2.5){$x$}
\put(3.3,3.8){$1/4$}
\put(3.3,3.1){$1/4$}
\put(3.3,2.6){$1/4$}
\put(3.3,2.1){$1/4$}

\put(4.1,3.8){$1$}
\put(4.1,3.1){$3$}
\put(4.1,2.5){$7$}
\put(4.1,1.7){$9$}

\put(6.5, 2.5){\vector(1,1){1.5}}
\put(6.5, 2.5){\vector(2,1){1.5}}
\put(6.5, 2.5){\vector(1,0){1.5}}
\put(6.5, 2.5){\vector(2,-1){1.5}}

\put(6.3,2.5){$y$}
\put(7.3,3.8){$1/4$}
\put(7.3,3.1){$1/4$}
\put(7.3,2.6){$1/4$}
\put(7.3,2.1){$1/4$}

\put(8.1,3.8){$2$}
\put(8.1,3.1){$4$}
\put(8.1,2.5){$6$}
\put(8.1,1.7){$8$}
\end{picture}
\end{figure}

As an example, let us assume that $x$ and $y$ are two lotteries with four different equally plausible states of the world each, as represented in Figure \ref{lottery}. The corresponding cumulative distribution functions $F_x$ and $F_y$ are graphically represented in Figure \ref{cdf1}. The EU maximizer will select lottery $y$ to lottery $x$, since the two lotteries have the same mean but $x$ has a wider spread around the mean. Instead, different $\tau$-maximizers will choose a different lottery depending on the quantile they focus on. For $\tau$=0, $y$ will be preferred over $x$ because the highest outcome in the worse-case scenario belongs to $y$. This is true for all values of $\tau \in(0,1/2)$. For $\tau\geq 1/2$, the preferred lottery becomes $x$ because, with a probability equal or larger than $1/2$, the minimum outcome that individuals can realize under $x$ is always higher than the minimum outcome which they can realize under $y$.

In our model, the two lotteries correspond to the prospects of migrating and non migrating. We claim that the different risk-aversion gradient of individuals results in different migration choices because they concentrate on a different quantile to take their decision. If $x$ is the prospect of migrating and $y$ is the prospect of staying, only relatively more risk-tolerant individuals would leave their homes.\footnote{In this simplified approach we are implicitly assuming that all individuals in the population face the same prospects from staying and from leaving. A different approach would instead predict the set of possibilities for any one individual, similarly to what it is done in vulnerability to poverty literature (\citealp{Chauduri_2002}).}

Note that, in the QM framework, individuals can be ordered with respect to their degree of risk-aversion by looking at the quantile chosen to make decisions among different prospects. In particular, as noted by \citet{Rostek_2010}, in ``the Quantile Maximization model, $\tau<\tau'$ if and only if a $\tau-$maximizer is weakly more averse toward downside risk than a $\tau'-$maximizer'' (\citealp{Rostek_2010}, p.354).

To conclude, the advantages of QM over EU can be summarized as follows: (i) QM does not require any hypothesis on the shape of individual utility functions, only that more (income or expenditure) is better than less; (ii) QM does not require full information on the entire support of distribution functions of future prospects, only on part of the domain (even just the extreme realizations, without any probabilistic information, in the extreme cases of maxmin and maxmax decision makers); (iii) QM allows to model a type of risk-aversion, namely, aversion to downside risks, which appears to be better suited to model decision under conflict; (iv) QM does not require a learning process based on past experiences similar to recent EU types of models proposed in the context of economic migration and is therefore more suitable for choices made in a forced displacement context.  

\section{Empirical strategy}

Our empirical strategy is structured in four steps: 1) Estimation of the two distributions of outcomes (lotteries) for people living under conflict who decide to stay or leave; 2) Predictions of migration choices for people with different risk preferences (focusing on different quantiles of the distributions) based on the QM model; 3) Test whether the QM predictions match observed risk preferences among stayers and leavers; 4) Test for a potential confounding factor.

The first step of our empirical strategy is to determine the distribution of potential outcomes for people opting to stay or leave conflict-affected areas. Outcomes are measured in terms of household expenditure per capita assuming that people will compare their living conditions under the two scenarios when taking the decision to flee conflict. The outcomes for stayers are observed and the distribution of these outcomes can be derived directly from data. The potential gains (outcomes) of moving from conflict to non conflict areas are not observed and require estimation. In order to estimate these potential gains for households living in conflict areas, we first estimate a welfare equation for non-conflict areas as a log-linear equation as follows:

\begin{equation}\label{ols}
ln(y_{i,nc}) = \alpha + \beta_{nc} X_{i,nc} + \epsilon_{i}
\end{equation}

where $y$ is household expenditure per capita, $i=1,2,\dots,n$ are all households living in non-conflict affected areas ($nc$), and $X_{i,nc}$ is a matrix of household-specific characteristics. The estimated coefficients $\hat{\beta}_{nc}$ are then used to predict the expenditure levels that households living in conflict areas could potentially have if they migrated to non-conflict-affected areas:

\begin{equation}\label{pols}
\hat{ln(y_{i,c})}=\hat{\alpha} + \hat{\beta}_{nc} X_{i,c}
\end{equation}

We then use Duan's smearing re-transformation to obtain levels of expenditure starting from the predicted log expenditures (\citealp{Duan_1983}). 

This procedure allows us to obtain two comparable distributions of potential outcomes (lotteries) for people living under conflict if they choose to stay or to leave. The first distribution can be interpreted as the distribution of gains for people who decide to stay (\emph{stayers}) and the second distribution can be interpreted as the distribution of gains for people who decide to leave (\emph{leavers}).

As in all prediction models of this kind, selection on unobservables is possible but less likely than in models looking at mean outcomes for treated and untreated individuals. In our case, we are not comparing two groups (treated and non treated) but two lotteries for the same group. The two lotteries relate to the same people with the same observable and unobservable characteristics. There may be unobservable characteristics relevant to predict consumption that bias consumption estimates but they are irrelevant for the decision to flee conflict if they are not observed by the decision makers. The other assumption we make (common to all cross-imputation exercises) is that the Betas of the prediction equation are constant if one migrates or stays. This is a strong assumption of course and we could still miss on some unobservables that are important to people when deciding to flee conflict, they are observed by the decision makers but not captured by the data and are unrelated to risk aversion. One of the likely factors that fall in this category is having kin in safe areas, a factor we do not observe and that is likely to affect the decision to migrate. To address this question, we have added “language” to the prediction of consumption in the case of migration as a proxy for kin networks.

The second step consists of comparing the two distributions of gains to find the migration choice that the QM model would predict for people with different risk preferences, or at the extreme, maxmin (risk averse) and maxmax (risk tolerant) individuals. This is done by means of stochastic dominance analysis of first degree, i.e. by comparing the Cumulative Distribution Functions (CDFs) of the two lotteries. This analysis reveals the highest gain (dominance) between the two lotteries for each quantile of the probability distributions.\footnote{Recall that the \textit{y-axis} of CDFs built on probabilistic functions represent the sum of probabilities up to the corresponding outcome. This is why we referred to \textit{downside risks} when we described the model.} It can tell us, therefore, what the migration preference of people focusing on different quantiles would be. Or, in a world of only maxmin and maxmax individuals, it will predict the migration choice made by these two groups. 

The third step is to test whether the predictions of the QM model match the risk preferences observed among stayers and leavers. This is done using a direct question on risk preferences administered to respondents and studying the association between migration status and risk preferences as follows:

\begin{equation}\label{migration}
M_{i} = \alpha + \beta RA_{i} + \gamma X_{i} + \eta_{i}
\end{equation}

where $M(0,1)$ is the migration status of heads of households $i$, $RA$ is the observed risk aversion of heads of households, $X$ is a vector of household and individual characteristics and $\eta$ is the error term. If the predictions of the QM model for risk averse and risk tolerant individuals match the risk preferences observed among stayers and leavers, we consider our model successfully tested. Note that we expect $RA_i$ to be independent of $\eta_{i}$ because many of the known correlates of $RA_i$ are observed and present in $X_i$.   

The fourth and final step is to test for a possible confounding factor. As already discussed, one possible confounding factor of our results is that conflict could increase risk-tolerance as found by \citet{Maarten_2012} for Burundi and by \citet{Callen_2014} for Afghanistan. If this is the case, results could be explained solely by the fact that stayers and leavers are exposed to conflict for different lengths of time. To test for this confounding factor we carry out a Difference-in-Differences (DiD) test. We consider risk-aversion as an outcome, conflict as treatment and migration as a control and interact conflict and migration to obtain the DiD estimator as follows:

\begin{equation}\label{risk}
RA_{i}=\alpha + \beta C_{i} + \gamma M_{i} + \theta C_{i}M_{i} + \phi X_{i} + \pi_{i}
\end{equation}

Where $RA_{i}$ is risk-aversion, $C_{i}$ is a dummy variable taking value $1$ if the individual has  experienced conflict and $0$ otherwise; $M_{i}$ is a dummy variable taking value $1$ if the individual never migrated over the period taken into account and $0$ otherwise, $X_i$ is a matrix of individual's socio-economic characteristics, and $\pi$ is the error term.

\section{Data and context}
The paper uses household and conflict data from Nigeria, the largest country in Africa in terms of population and GDP and also one of the largest countries in terms of IDPs with over 3.3 m in 2023. Nigeria has endured high degrees of violence in many parts of the country for decades but the most intense and prolonged internal conflict has been the one generated by the Boko Haram insurgency in the North-East of the country over the past twenty years. This is the area of the country we study covering the period that saw the most intense degree of violence, between 2010 and 2016. 

The household data we use is the General Household Survey (GHS) administered by the National Bureau of Statistics (NBS) of the Federal Government of Nigeria in collaboration with the World Bank Living Standards Measurement Study.\footnote{All data and ancillary files can be found in the World Bank microdata library. For an example of the latest GHS survey see: http://microdata.worldbank.org/index.php/catalog/2734.} Since 2010, this survey contains a panel component of 5,000 households administered during the post-planting and post-harvest seasons every other year, which is what we use in this paper. Each of these two components is a self-contained nationally and regionally representative survey and is administered in a different calendar year. This makes a total of six points in time with one point in time per calendar year between 2010 and 2016 (in 2015 the survey was not administered due to the degree of violence in the country). This is the period we consider, which captures the evolution of the most intense part of the Boko Haram conflict with its peak around the end of 2014. 

The panel is a fixed (non-rotating) panel where missing households are traced to their new addresses and interviewed when possible. This reduced the attrition rate at the end of the six rounds to 8.3 percent of households. The survey records the reason why households do not respond, as summarized in Table \ref{Table_Attrition_Reason} for the case of the last visit in 2016. A third of the attrition is due to the impossibility of interviewing households living in crisis areas. In particular, 14 enumeration areas could not be visited in Borno and Yobe States at the peak of the conflict in 2015, which is potentially problematic for our analysis. The original sample weights are adjusted by the national statical agency to account for attrition using household nonresponse rates by enumeration areas  (we refer the reader to the Basic Information Document, p.21 and its Appendix 5 for further details, \citealp{GHS_2016}). To test whether this correction restores the means of the original sample, we conducted means tests with the 2010 sample on all variables of interest between the full sample and the sample with attrition.\footnote{All covariates used in the paper are described in Table \ref{Variables}.} Table \ref{Variables} describes all relevant variables we use in the paper and Table \ref{T_test} reports results of the means tests. The test shows no significant differences across all variables of interest with the exception of household size and the marital status of the head of the household. 

\begin{figure}[!htb]
\caption{Cumulative distribution functions of predicted values for per capita expenditure, first visit}\label{Per_capita_expenditure_dominance}
\begin{center}
\includegraphics[scale=0.45]{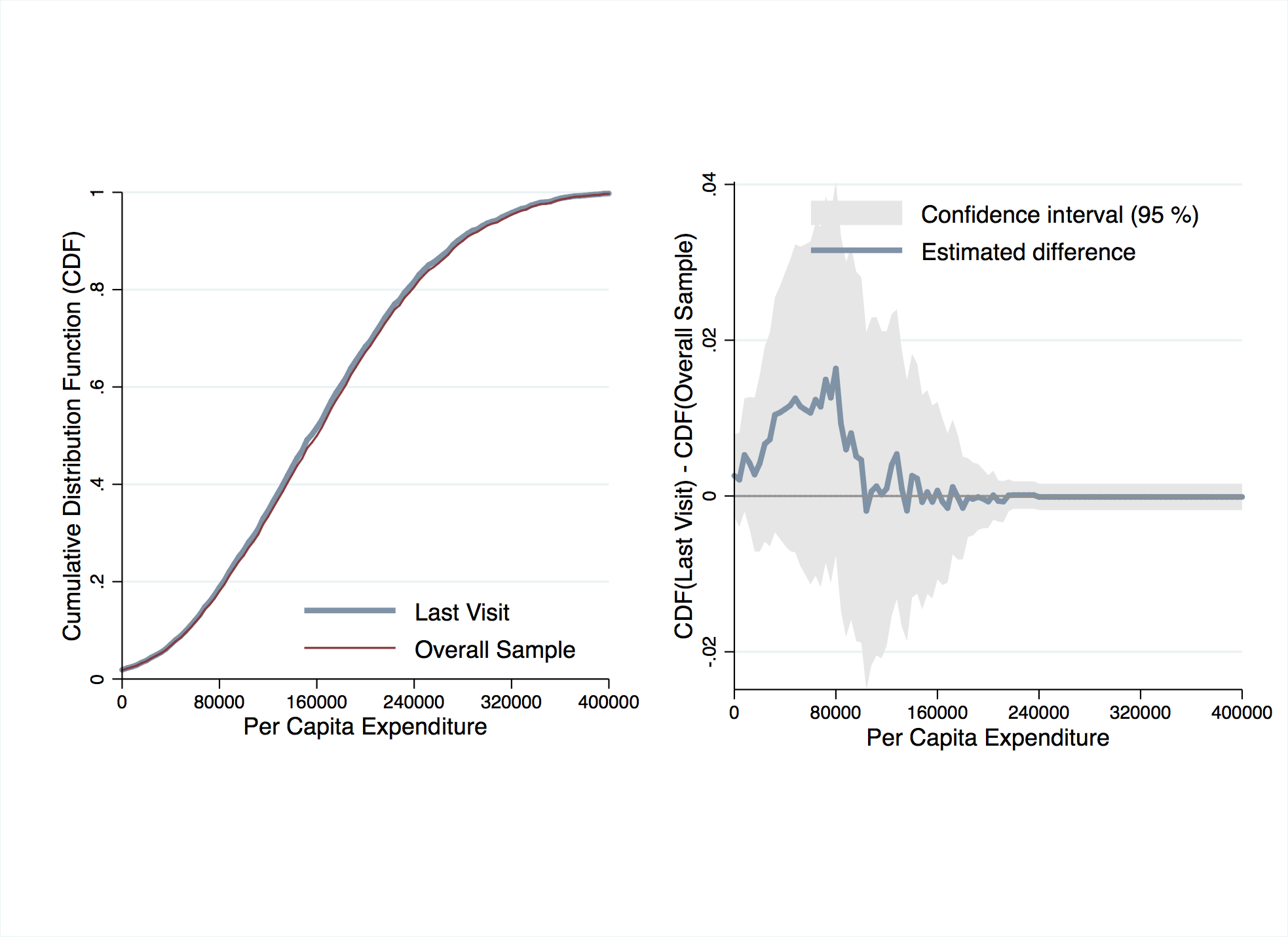}
Source: Authors' estimations based on GHS Data
\end{center}
\end{figure}

\begin{figure}[!htb]
\caption{Cumulative distribution functions of predicted values for risk-tolerance}\label{Risk_love_dominance}
\begin{center}
\includegraphics[scale=0.45]{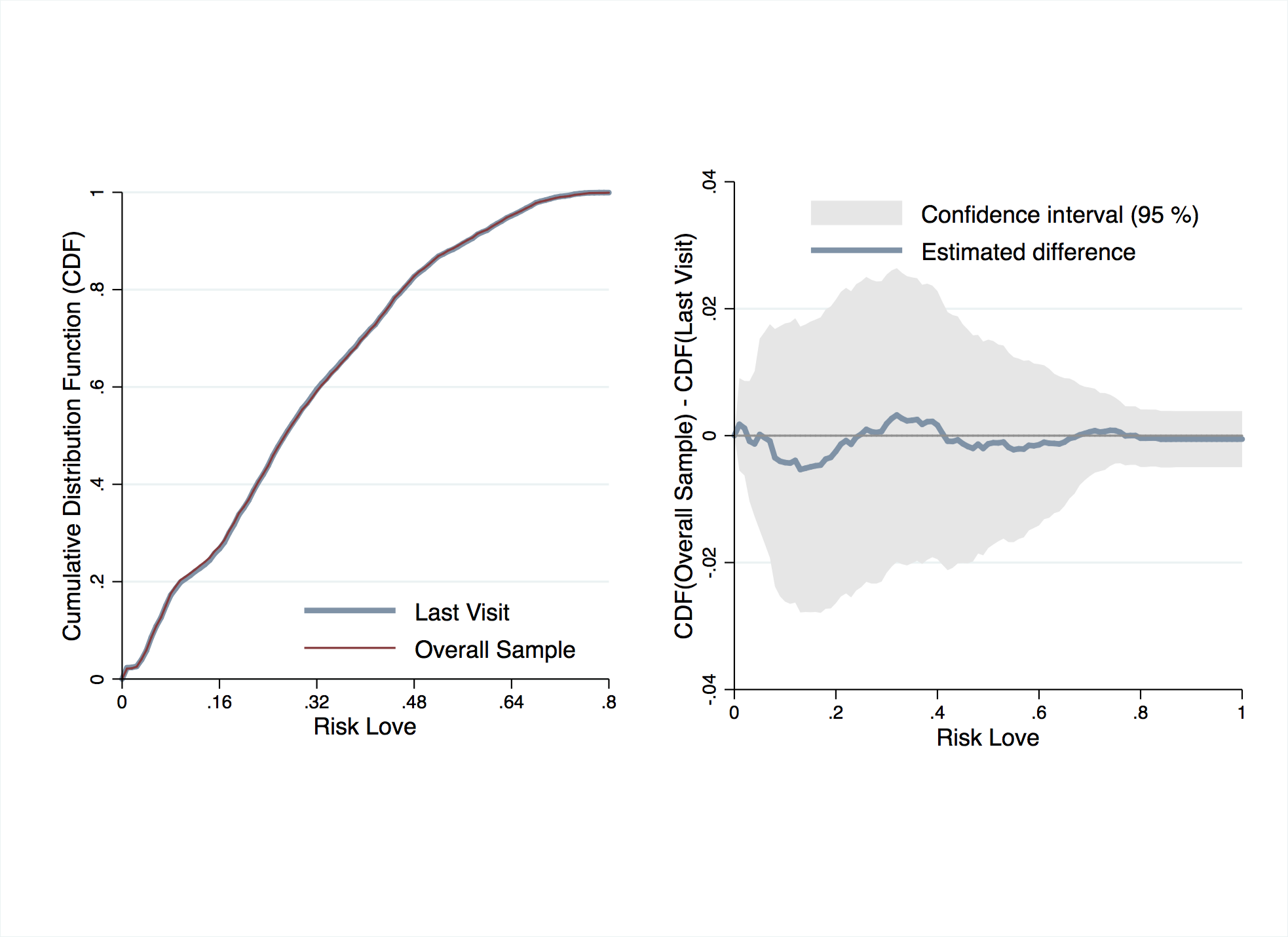}
Source: Authors' estimations based on GHS Data
\end{center}
\end{figure}

However, attrition could affect the prediction capacity of these variables when used together in a multivariate form in models that include our key variables: expenditure per capita and risk aversion. The variables considered in Table \ref{T_test} enter the models used to predict expenditure (equations \ref{ols} and  \ref{pols}), and to study the association between migration and risk aversion (equation \ref{migration}) in a multivariate form. To test for non random attrition  on per capita expenditure and risk-aversion respectively, we compare the Cumulative Distribution Functions (CDFs) of the predicted values of these variables (based on the predictors listed in Table \ref{T_test}) and test for stochastic dominance of first degree between the samples with and without attrition. This is a rather simple but effective test. We want to make sure that predictors are stable with respect to these variables whether we have attrition or not. If the predictions generated by the sample with attrition are different from those generated by the sample without attrition, our results would be biased. For per capita expenditure, we observe this variable all along the panel. We can therefore use the 2010 full sample to make predictions and then compare the two distributions of predicted values for the sample with and without attrition. Results are shown in Table \ref{pcexpdr_attrition} and Figure \ref{Per_capita_expenditure_dominance} and show that there is no difference in the distributions of predicted values between the two samples. For risk-aversion, we observe this variable only in 2016. We assume that risk-aversion is time-invariant and we use individual characteristics in 2010 to predict the probability of being risk-tolerant for the full panel based on individuals left in the 2016 panel sample with attrition. Results are shown in Table \ref{risk_love_logit} and Figure \ref{Risk_love_dominance} and show again that there is no difference in the distribution of predicted values of risk-aversion between the overall sample and the sample with attrition. We can therefore conclude that attrition does not bias our results with respect to per capita expenditure and risk-aversion.

The migration status of households is measured across the 774 Local Government Areas (LGAs), the lowest administrative subdivision available in our data. We consider a household to have migrated if it changed LGA in any of the years considered. We consider three migration statuses for Equation \ref{migration}. People who have migrated from areas with no conflict, people who have migrated from areas with some conflict (conflict in at least one year of the six years considered), and people who have migrated from areas with permanent conflict (conflict in all six years considered). Overall, we have 329 migrant households (6.6 percent of the sample) and 90.5 percent of these migrants migrated within one of the six geographical regions considered. Therefore, most of the migration observed is short range migration.

To control for possible predictors of risk-aversion in equation \ref{migration}, we include most known correlates of risk preferences. Risk-aversion has been found to be higher in women and lower in men or individuals with higher levels of testosterone (\citealp{Eckel_2008}; \citealp{Apicella_2008}; \citealp{Johnson_2006}), and higher for married and older individuals (Kuznets, 1964; \citealp{Myrdal_1968}); it is associated with lower cognitive abilities (\citealp{Dohmen_2010}), lower levels of skills, occupational status and income (\citealp{Hartog_2002}). Risk preferences were also found to be genetic and transmissible through generations (\citealp{Dohmen_2011}; \citealp{Becker_2016}). Our models control for age, education, marital status and occupational status. We do not have information on testosterone levels, genetic heritage or cognitive abilities but it can be argued that gender is a good proxy for testosterone levels whereas education and employment status are good proxies for cognitive abilities. We also measure migration across small administrative areas where genetic diversity is not expected to be very large. Instead, income (proxied by expenditure in our data) is our measure of outcomes and should not be controlled for in the migration equation by design.  

The 2016 GHS also contains a battery of behavioral questions, which include questions on risk-aversion. To assess the degree of risk-aversion we used the following question: \textit{Suppose you want to invest some money. Which option do you prefer? A) Investing in a business where I can't lose money but has low profits; B) Investing in a business where there is a chance I can lose money but potentially brings high profits.} Empirical studies on migration and risk-aversion typically use as measures of risk-aversion a self-assessment measured on a scale from 0 to 10 (\citealp{Bonin_2009}; \citealp{Jaeger_2010}; \citealp{Akguc_2016}). These studies then argue that this scale is a proper measure of risk-aversion because it is closely associated with direct measurements of risk-aversion as shown by \citet{Jaeger_2010} and \citet{Dohmen_2011}. In our study, we do not need this extra verification as we have a question that measures risk preferences \textit{directly}. 

\begin{center}

\begin{figure}[!htb]
\caption{Incidence of Fatalities by Local Government Areas, 2011-2016}\label{Acled}
\includegraphics[scale=1]{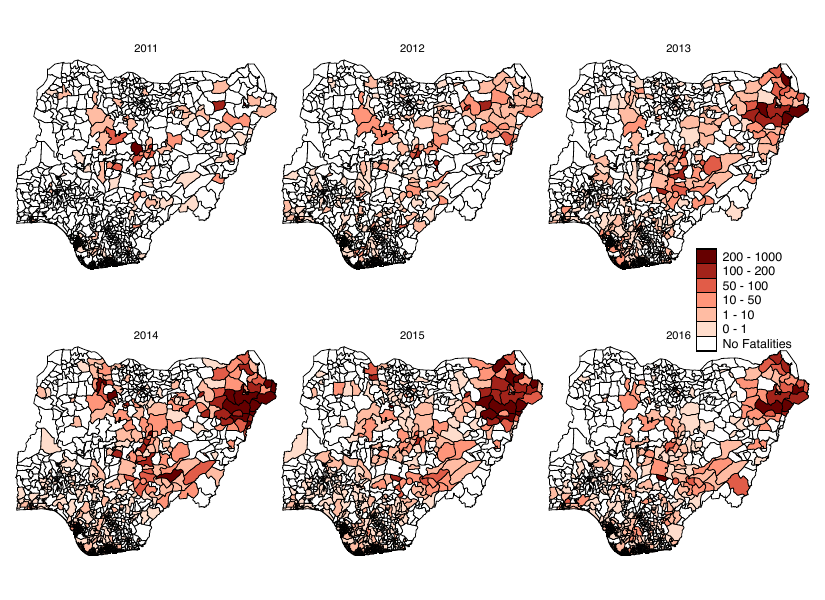}

Source: Authors' elaboration based on ACLED Data (http://www.acleddata.com/data/)
\end{figure}
\end{center}

The second data set we use is the the Armed Conflict Location \& Event Data (ACLED). These data cover daily conflicts episodes in the country since 1997 and include information on the cause of conflict, perpetrators, casualties and fatalities. We will use the number of fatalities by LGA and match these data with the GHS panel data for the period 2010-2016. We then subdivide LGAs in areas that had no conflict (in any of the years considered), areas with some conflict (defined as areas with fatalities due to conflict in at least one of the years considered but less than six years) and areas always in conflict (defined as areas with fatalities due to conflict in all six years considered).\footnote{Data are available on-line and include codebooks, user guides, questionnaires and other relevant material (see http://www.acleddata.com/data/).} The ACLED data set is not perfect as many episodes of violence are not reported or detected but the map of casualties constructed with these data shows correctly all the areas where the conflict was known to be intense (see Figure \ref{Acled}). 

\section{Results}

We report results following the order described in the empirical strategy. As a first step we estimate the distributions of outcomes for people living in conflict areas. Table \ref{prediction} reports coefficients and standard errors of the predicted model for households living in non-conflict areas (Equation \ref{ols}).\footnote{Note that the equation in Table \ref{prediction} was also run without variables related to the quality of dwelling. This was to allow for the hypothesis that people may expect to live in makeshift shelters once they move. Results of this test showed that the R squared decreased by 4 percetage points whereas the following results on the dominance analysis remained intact. We also know that most migrants captured by the survey moved to accomodations not too dissimilar from their own homes with the exception of some of the IDPs hosted in camps. Moreover, migrants are generally expected to overestimate rather than underestimate the benefits of migration.} Next, the estimated coefficients $\hat{\beta}_{nc}$ are used to predict the expenditure levels that households living in conflict areas could potentially have if they migrated to non-conflict-affected areas (Equation \ref{pols}). We then use Duan's smearing re-transformation to obtain levels of expenditure starting from the predicted log expenditures (\citealp{Duan_1983}).

\begin{figure}[!htb]
\caption{Cumulative distribution functions of households living in conflict areas}\label{dominance}
\begin{center}
\includegraphics[scale=0.4]{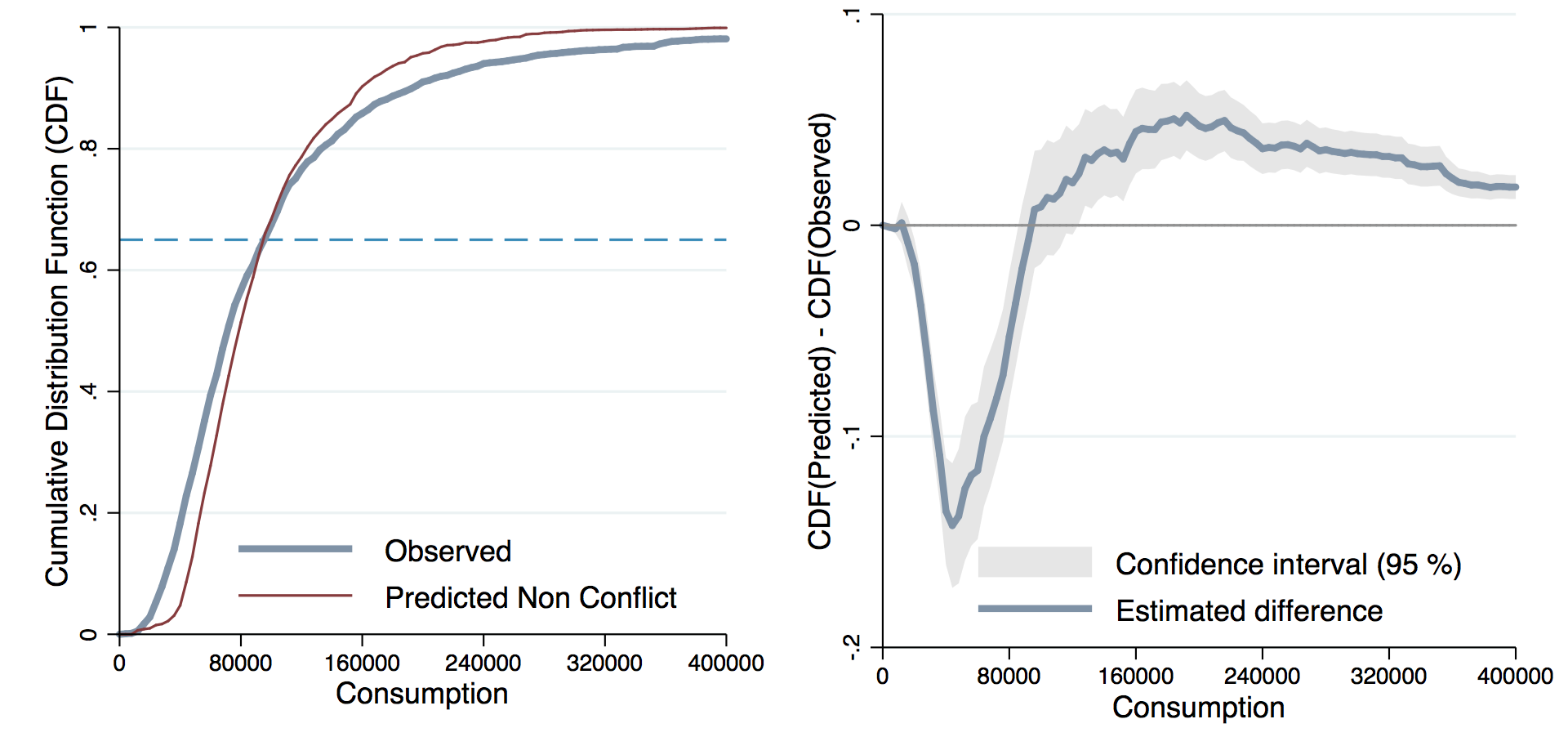}
Source: Authors' estimations based on GHS and ACLED Data
\end{center}
\end{figure}

As a second step, the observed distribution of expenditure per capita of people living in conflict areas is compared with the distribution of predicted values for the same people if they migrated to non-conflict areas. Figure \ref{dominance} shows the results using the 2016 GHS data. It is clear that the two curves cross in one point splitting the samples into two parts. The left end side of the CDFs shows that the curve of predicted values (gains from leaving) dominates the curve of observed values (gains from staying). The QM model would therefore predict that maxmin (risk averse) individuals who focus on the left hand side of the distributions (low gains) would opt to leave. Instead, the opposite is true for the right end side of the CDFs. In this case, the QM model would predict that maxmax (risk tolerant) individuals who focus on the higher end of the distributions (high gains) would opt to stay. As shown in the right hand panel of Figure \ref{dominance}, these differences are statistically significant. Therefore, the QM model predicts maxmin (risk-averse) individuals to leave and maxmax (risk-tolerant) individuals to stay.   

As a third step, we test whether these QM predictions match reality. We test whether risk-averse (maxmin) individuals  are effectively associated with migrating and risk-tolerant (maxmax) individuals are effectively associated with staying. Recall that migrants are defined as those households who have changed LGA at some point in time during the six panel waves and that risk-aversion is assessed using a question asked in a module of the 2016 GHS data as described in the data section. We test the association between migration and risk-tolerance for household heads controlling for a set of household socio-economic characteristics. 

Results are shown in Table \ref{prediction_risk}.\footnote{Note that some of the covariates related to the quality of dwelling are dropped in Table \ref{prediction_risk} because of perfect prediction. The variable ``language'' is also not included in this table and the next because it almost perfectly predicts areas in conflict. For instance, people living in areas affected by the Boko Haram conflict speak almost invariably Kanun or Hausa whereas people living in violent areas in the South speak mostly Igbo. As one would expect, language is closely associated with location.} The association between migration and risk-tolerance is negative and significant in all three specifications of the equation. The three specifications differ in the dependent variable. In all three equations, ``0'' corresponds to migrants and ``1'' corresponds to non-migrants. However, the ``1'' changes in the three equations to capture different degrees of conflict in the area of origin. Equation (1) captures non-migrants irrespective of whether the area of origin had conflict in previous periods (2010-2014) or not. Equation (2) uses a sub-sample of non-migrants who originates from areas that had some conflict over the previous periods, and Equation (3) reduces the sample to people in areas that had always conflict over the previous periods. Results show that the coefficient is always significant and that the size of the coefficient increases from Equation (1) to Equation (3). This shows that those who opt to migrate are the maxmin (risk averse) individuals and those who stay are  the maxmax (risk tolerant) individuals as predicted by the QM model with the dominance analysis. It also shows that the positive association between staying and risk-tolerance is stronger for people who sustained longer spells of violence. 

As a fourth and last step, we test for a possible confounding factor. Recall that conflict could increase risk-tolerance as found elsewhere and that the lower risk-aversion that we find among stayers in Table \ref{prediction_risk} could be explained solely by the fact that stayers have experienced a longer period of violence as compared to leavers. To test for this confounding factor we carry out a Difference-in-Differences (DiD) test as described in the empirical strategy section. 

Results are reported in Table \ref{prediction_did} and show that conflict and the interaction between these two variables (the DiD coefficient) are both non significant. Note also that the low R-squared and the insignificance of the independent variables are further evidence that the covariates of risk-aversion in Table \ref{prediction_risk} are independent. We can therefore reject the hypothesis that conflict affects risk preferences and explains the difference in risk preferences between movers and stayers.  

\section{Conclusion}
The paper departed from standard utility theory under uncertainty positing that individuals living in conflict areas are quantile maximizers as opposed to expected utility maximizers. In its most extreme form, quantile maximization implies that individuals tend to cluster into two groups, maxmin (risk-averse) and maxmax (risk-tolerant) individuals. 

We then proposed a four-step approach to test the model with empirical data: 1) Estimation of the distributions of outcomes (lotteries) for people living under conflict who decide to stay or leave; 2) Predictions of migration choices for people with different risk preferences (focusing on different quantiles of the distributions) based on the QM model; 3) Test whether the QM predictions match observed risk preferences among stayers and leavers; 4) Test for a potential confounding factor.

Using household and conflict data from Nigeria, we found that the predictions of the QM model correctly identify stayers and leavers based on risk preferences. The CDFs of observed and predicted gains from staying and migrating crossed in one point. We also found that risk-averse individuals have a greater propensity to migrate whereas risk-tolerant individuals have a greater propensity to stay as predicted by the dominance analysis. This association is very significant and increases passing from general migration to migration from high conflict areas. We also tested for a possible confounding factor related to the possibility that conflict affects risk preferences and explains the observed difference between migrants and non-migrants. We used a DiD estimator to test for this possibility and found no evidence that conflict affects risk-aversion.

How do these findings fare with the existing literature? In our knowledge, this is the first paper of its kind. We are not aware of another paper that attempted to use a similar model of decision making under conflict and then test the model empirically with household data. The closest work to our study is an unpublished experiment conducted on combatants and non combatants in conflict and non conflict zones of Aleppo in Syria in 2013 and 2014 (Whitt and Mironova, 2017). Using a variation of the \citeauthor{Eckel_2002} (2002, 2008) risk game, the authors find combatants and non-combatants in rebel-held conflict zones to be more risk-tolerant than other Syrians in other parts of the country who fled abroad to Turkey. Therefore, the only other similar study we found would confirm our results on Nigeria. 

Instead, these findings are in stark contrast with the literature on economic migration and risk preferences (\citealp{Jaeger_2010} and \citealp{Akguc_2016}). There is a remarkable difference between risk-tolerant economic migrants and risk-averse forced migrants. The cause of flight (economic or flight from violence) is therefore a discriminant in the self-selection process of migrant populations. Forced migration in Nigeria seems to be driven by risk-aversion and a desire to protect one's own minimum living standards as opposed to leaving in search of greater opportunities and rewards. 

These findings have important policy implications for the place of origin of migrants. Stopping or reducing conflict would be a necessary condition for most of the forced migrants to return home and, possibly, a sufficient condition if the place of origin guarantees basic needs. Indeed, most of the forced migrants tend to settle close to the place of origin in the expectation of going home when peace is restored. Evidence on economic migrants shows instead that this group is more likely to focus on growth opportunities and is ready to travel far and stay away for extended periods of time irrespective of conflict. This group would require substantial improvements in the place of origin to justify return, a condition that is rarely met in the short or medium term in a development context.

These same findings have also important policy implications for the place of destination of migrants. Risk-averse forced migrants who tend to settle close to the place of origin and expect to return when conditions allow would be less inclined to take entrepreneurial risks and invest in the place of destination whereas they are more likely to endure subsistence types of livelihoods. Instead, economic migrants who are more risk tolerant and seek long-term opportunities in better off places would be less likely to endure subsistence conditions and more likely to travel to a further destination even if the journeys imply important risks. This has implications for the type of policies that host governments may want to adopt for economic migrants as opposed to forced migrants. When compared with results on economic migration, our results suggest that bundling economic and forced migrants into one class does not serve well neither the place of origin nor the place of destination.  
    
\newpage

\bibliographystyle{plainnat}
\bibliography{Ceriani_Verme_Bib}


\newpage

\section*{List of Figures}

\newpage

\setlength{\unitlength}{1.2cm}

\begin{figure}[!htb]
\caption{Cumulative Distribution Function of $x$(blue) and $y$(red)}\label{cdf1}
\begin{center}
\begin{picture}(6,6)(0,0)
\put(0.5,0){\vector(0,1){5.5}}
\put(0.5,0){\vector(1,0){5.5}}
\linethickness{0.8mm}
\put(1,0.1){\textcolor{myblue}{\line(-1,0){0.5}}}
\put(2,1.35){\textcolor{myblue}{\line(-1,0){1}}}
\put(4,2.5){\textcolor{myblue}{\line(-1,0){2}}}
\put(5,3.75){\textcolor{myblue}{\line(-1,0){1}}}
\put(5,5){\textcolor{myblue}{\line(1,0){0.5}}}

\put(1,1.35){\textcolor{myblue}{\circle*{0.15}}}
\put(2,2.5){\textcolor{myblue}{\circle*{0.15}}}
\put(4,3.75){\textcolor{myblue}{\circle*{0.15}}}
\put(5,5){\textcolor{myblue}{\circle*{0.15}}}

\put(1,0.1){\textcolor{myblue}{\circle{0.15}}}
\put(2,1.35){\textcolor{myblue}{\circle{0.15}}}
\put(4,2.5){\textcolor{myblue}{\circle{0.15}}}
\put(5,3.75){\textcolor{myblue}{\circle{0.15}}}

\multiput(1,1.35)(0,-0.1){12}{\textcolor{myblue}{\line(0,-1){0.05}}}
\multiput(2,2.5)(0,-0.1){12}{\textcolor{myblue}{\line(0,-1){0.05}}}
\multiput(4,3.75)(0,-0.1){12}{\textcolor{myblue}{\line(0,-1){0.05}}}
\multiput(5,5)(0,-0.1){12}{\textcolor{myblue}{\line(0,-1){0.05}}}

\put(1.5,0){\textcolor{red}{\line(-1,0){1}}}
\put(2.5,1.25){\textcolor{red}{\line(-1,0){1}}}
\put(3.5,2.60){\textcolor{red}{\line(-1,0){1}}}
\put(4.5,3.85){\textcolor{red}{\line(-1,0){1}}}
\put(4.5,5.1){\textcolor{red}{\line(1,0){1}}}

\put(1.5,1.25){\textcolor{red}{\circle*{0.15}}}
\put(2.5,2.60){\textcolor{red}{\circle*{0.15}}}
\put(3.5,3.85){\textcolor{red}{\circle*{0.15}}}
\put(4.5,5.1){\textcolor{red}{\circle*{0.15}}}

\put(1.5,0){\textcolor{red}{\circle{0.15}}}
\put(2.5,1.25){\textcolor{red}{\circle{0.15}}}
\put(3.5,2.60){\textcolor{red}{\circle{0.15}}}
\put(4.5,3.85){\textcolor{red}{\circle{0.15}}}

\multiput(1.5,1.25)(0,-0.1){13}{\textcolor{red}{\line(0,-1){0.05}}}
\multiput(2.5,2.60)(0,-0.1){13}{\textcolor{red}{\line(0,-1){0.05}}}
\multiput(3.5,3.85)(0,-0.1){12}{\textcolor{red}{\line(0,-1){0.05}}}
\multiput(4.5,5.1)(0,-0.1){12}{\textcolor{red}{\line(0,-1){0.05}}}

\thinlines

\multiput(1,-0.15)(0.5,0){10}{\line(0,1){0.25}}
\put(0.3,-0.3){0}
\put(0.9,-0.5){1}
\put(1.4,-0.5){2}
\put(1.9,-0.5){3}
\put(2.4,-0.5){4}
\put(3.4,-0.5){6}
\put(3.9,-0.5){7}
\put(4.4,-0.5){8}
\put(4.9,-0.5){9}

\multiput(0.35,1.25)(0,1.25){4}{\line(1,0){0.25}}
\put(0.15,1.25){$\frac{1}{4}$}
\put(0.15,2.5){$\frac{1}{2}$}
\put(0.15,3.75){$\frac{3}{4}$}
\put(0.15,4.9){$1$}
\end{picture}
\end{center}
\end{figure}
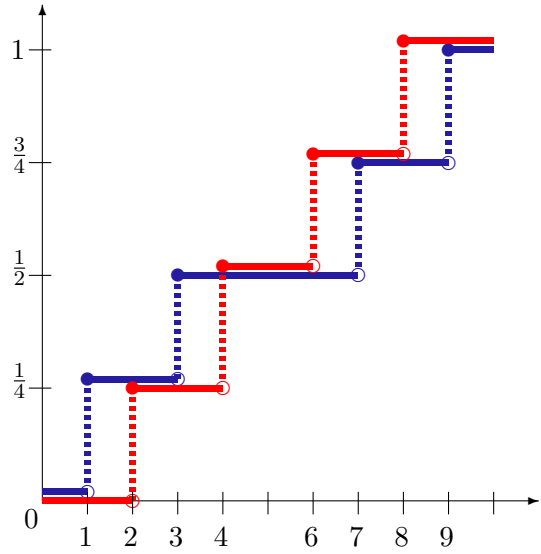

\newpage

\newpage

\newpage

\newpage

\newpage

\section*{List of Tables}

\begin{table}[!h] 
\caption{Household interview status during the last visit}\label{Table_Attrition_Reason}
\begin{center}
\begin{tabular}{lrr}
\hline
& Observations & Frequency\\
\hline 
Interviewed (in main survey phase) & 4,534 & 90.68\\
Tracked (and interviewd) & 48 & 0.96\\
Refused & 47 & 0.94\\
Not found & 100 & 2.00\\
Dead & 84 & 1.68\\
Moved away (not tracked) & 48 & 0.96\\
Crisis area & 139 & 2.78\\
\hline 
Total & 5,000 & 100 \\
Total number of migrants & 329 & 6.6 \\
\hline
\end{tabular}
\end{center}
Source: Authors' estimations based on GHS Data
\end{table}

\begin{sidewaystable}
\caption{Description of variables used in Equation \ref{ols}}\label{Variables}
\centering
\begin{tabular}{p{3cm}p{5cm} p{9cm}}
\textbf{Variable} & \textbf{label} & \textbf{Description}\\
\hline
ln\_exp & log expenditure & Per capita Household Expenditure expressed in Logaritm. This variable is the expenditure aggregate constructed by the World Bank for the purpose of poverty measurement\\
\hline 
hhsize & Household Size & Number of Household Members living in the household at the moment of the interview\\
\hline
demo\_sh\_depentants & Share of Dependent Members & Dependent household members, defined as individuals aged less than 15 and over 64 living in the household, as a share of total household members\\
\hline
state & State & Categorical variable where each level is one of the 36 Nigerian States plus the Federal Capital Territory of Abuja: Abia; Adamawa; Akwa Ibom; Anambra; Bauchi; Bayelsa; Benue; Borno; Cross-rivers; Delta; Ebonyi; Edo; Ekiti; Enugu; Gombe; Imo; Jigawa; Kaduna; Kano; Katsina; Kebbi; Kogi; Kwara; Lagos; Nassarawa; Niger; Ogun; Ondo; Osun; Oyo; Plateau; Rivers; Sokoto; Taraba; Yobe; Zamfara; FCT Abuja
 \\
\hline
lga & LGA & Categorical variable where each level is one Local Government Areas. Of the total 774 LGAs Nigeria is divided into, the dataset contains observations from 467 LGAs.\\
\hline
rururb & Type of Settlement & Type of settlement (1=rural; 0=urban)\\
\hline
hh\_sex & HH Gender & Gender of the Household Head (1=male; 0=female)\\
\hline
hh\_agey & HH Age & Age of the household head\\
\hline
hh\_eduyrs & HH Year of Education & Number of years of education attained by the household head\\
\hline
hh\_empl & HH Employed & Status in employment of the household head (1=employed; 0=unemployed or inactive)\\
\hline
hh\_marstat & HH Married & Marital Status of the household head (1=married; 0=Divorced, separated, widowed, or never married)\\
\hline
hh\_language & HH Principal Language & Principal language spoken in the household (1=Hausa; 2=Igbo; 3=Yoruba; 4=Other)\\
\end{tabular}
\end{sidewaystable}

\begin{sidewaystable}
\centering
Table 2: Description of variables used in Equation \ref{ols} (cont.)
\begin{tabular}{p{3cm}p{5cm} p{9cm}}
\textbf{Variable} & \textbf{Label} & \textbf{Description}\\
\hline
dwel\_rooms & Number of Rooms & Number of habitable rooms in the dwelling\\
\hline
dwel\_roof  & Quality of Roof & Quality of the material used for the dwelling's roof (1=Low quality, 1=Medium quality , 3=High quality)\\
\hline
dwel\_wall &  Quality of Wall & Quality of the material used for the dwelling's walls (1=Low quality, 1=Medium quality , 3=High quality)\\
\hline
dwel\_floor  &  Quality of Floor & Quality of the material used for the dwelling's floors (1=Low quality, 1=Medium quality , 3=High quality)\\
\hline
dwel\_toilet & Type of Toilet & Type of main toilet facility (1=No facility,  2=Flush toilet, 3=Improved pit latrine, 4=Uncovered pit latrine, 5=Other)
\\
\hline
dwel\_fuellight & Main source of Lighting & Main source of lighting (1=Electricity/gas, 2=Kerosene, 3=Candles/battery, 4=Other) \\
\hline
dwel\_fuelcook & Main heat source for cooking& Main heat source for cooking (1=Firewood, 2=Charcoal, 3=Kerosene/oil, 4=Electricity/gas, 5=Other) \\
\hline
dwel\_gdisp & Type of garbage disposal & Type of garbage disposal (1=In bins collected by public service; 2= Disposal in compound; 3=Discarded in empty lots/streets/rivers)\\
\hline
own\_tv & TV & Household owns a TV\\
\hline
own\_fridge & Fridge &  Household owns a Refrigerator\\
\hline
own\_stove & Stove & Household owns a Stove\\
\hline
own\_bcycle & Bicycle & Household owns a Bicycle\\
\hline
own\_car & Car & Household own a Car\\
\hline
own\_iron & Iron & Household own an Iron\\
\hline
conflict & Conflict & Dummy variable equal to 1 if in the Local Government Area at the time of GHS survey there have been fatalities due to conflict events as reported in the Acled Dataset\\
\hline
\end{tabular}
\end{sidewaystable}

\newpage
\begin{center}
\begin{longtable}[!h]{lcccc}
\caption{Descriptive Statistics}\label{T_test}
\\
 \hline
	&	Sample with  	&	 Overall	&	 F	&	 Prob $>$ F\\
	&	 Attrition 	&	 Sample	&	 	&	 \\
\hline
\multicolumn{5}{l}{\textbf{Social and Economic Variables}}\\
Household Size	&	5.491	&	5.365	&	3.361	&	0.067	\\
Share of Dependent Members	&	0.432	&	0.433	&	0.038	&	0.845	\\
Type of Settlement (Rural==1)	&	0.400	&	0.409	&	0.535	&	0.465	\\
HH Gender (1=Male)	&	0.851	&	0.843	&	0.888	&	0.346	\\
HH Age	&	49.578	&	49.602	&	0.004	&	0.947	\\
HH Years of Education	&	6.545	&	6.542	&	0.000	&	0.985	\\
HH Employed	&	0.916	&	0.910	&	0.740	&	0.390	\\
HH Married	&	0.805	&	0.790	&	2.520	&	0.112	\\
HH Gender * HH Married	&	0.788	&	0.771	&	2.613	&	0.106	\\
HH Language	&	           .	&	           .	&	           .	&	           .	\\
\multicolumn{1}{r}{Hausa}	&	0.280	&	0.267	&	1.383	&	0.240	\\
\multicolumn{1}{r}{Igbo}	&	0.174	&	0.173	&	0.028	&	0.867	\\
\multicolumn{1}{r}{Yoruba}	&	0.238	&	0.252	&	1.549	&	0.213	\\
\multicolumn{1}{r}{Other}	&	0.308	&	0.308	&	0.002	&	0.966	\\
\multicolumn{5}{l}{\textbf{Quality of Dwelling}}\\									
Number of Rooms	&	3.440	&	3.386	&	1.140	&	0.286	\\
Quality of Roof	&	           .	&	           .	&	           .	&	           .	\\
\multicolumn{1}{r}{Low}	&	0.176	&	0.184	&	0.781	&	0.377	\\
\multicolumn{1}{r}{Medium}	&	0.808	&	0.799	&	0.867	&	0.352	\\
\multicolumn{1}{r}{High}	&	0.016	&	0.017	&	0.061	&	0.805	\\
Quality of Wall	&	           .	&	           .	&	           .	&	           .	\\
\multicolumn{1}{r}{Low}	&	0.415	&	0.417	&	0.016	&	0.901	\\
\multicolumn{1}{r}{Medium}	&	0.100	&	0.098	&	0.116	&	0.734	\\
\multicolumn{1}{r}{High}	&	0.485	&	0.486	&	0.006	&	0.938	\\
Quality of Floor	&	           .	&	           .	&	           .	&	           .	\\
\multicolumn{1}{r}{Low}	&	0.103	&	0.108	&	0.491	&	0.483	\\
\multicolumn{1}{r}{Medium}	&	0.191	&	0.189	&	0.078	&	0.780	\\
\multicolumn{1}{r}{High}	&	0.706	&	0.703	&	0.054	&	0.816	\\
Type of Toilet	&	           .	&	           .	&	           .	&	           .	\\
\multicolumn{1}{r}{No Facility}	&	0.184	&	0.185	&	0.026	&	0.871	\\
\multicolumn{1}{r}{Flush Toilet}	&	0.188	&	0.191	&	0.059	&	0.809	\\
\multicolumn{1}{r}{Improved Pit Latrine}	&	0.026	&	0.026	&	0.001	&	0.972	\\
\multicolumn{1}{r}{Uncovered Pit Latrine}	&	0.476	&	0.467	&	0.550	&	0.458	\\
\multicolumn{1}{r}{Other}	&	0.126	&	0.131	&	0.474	&	0.491	\\
Main Source of Lighting	&	           .	&	           .	&	           .	&	           .	\\
\multicolumn{1}{r}{Electricity/Gas}	&	0.353	&	0.349	&	0.118	&	0.731	\\
\multicolumn{1}{r}{Kerosene}	&	0.371	&	0.371	&	0.004	&	0.949	\\
\multicolumn{1}{r}{Candles/Batterty}	&	0.134	&	0.133	&	0.009	&	0.924	\\
\multicolumn{1}{r}{Other}	&	0.142	&	0.147	&	0.474	&	0.491	\\
Main heat source for cooking	&	           .	&	           .	&	           .	&	           .	\\
\multicolumn{1}{r}{Firewood}	&	0.659	&	0.645	&	1.346	&	0.246	\\
\multicolumn{1}{r}{Charcoal}	&	0.009	&	0.010	&	0.185	&	0.667	\\
\multicolumn{1}{r}{Kerosene/Oil}	&	0.288	&	0.293	&	0.182	&	0.670	\\
\multicolumn{1}{r}{Electricity/Gas}	&	0.016	&	0.017	&	0.105	&	0.745	\\
\multicolumn{1}{r}{Other}	&	0.029	&	0.036	&	3.023	&	0.082	\\
Type of Garbage disposal	&	           .	&	           .	&	           .	&	           .	\\
\multicolumn{1}{r}{Bins}	&	0.146	&	0.150	&	0.164	&	0.685	\\
\multicolumn{1}{r}{Compound}	&	0.329	&	0.317	&	1.385	&	0.239	\\
\multicolumn{1}{r}{Empty Lots/streets/rivers}	&	0.525	&	0.534	&	0.505	&	0.478	\\
\multicolumn{5}{l}{\textbf{Assets}}\\									
TV	&	0.428	&	0.427	&	0.012	&	0.915	\\
Fridge	&	0.195	&	0.194	&	0.012	&	0.912	\\
Stove	&	0.494	&	0.499	&	0.182	&	0.669	\\
Bicycle	&	0.191	&	0.187	&	0.163	&	0.686	\\
Car	&	0.098	&	0.097	&	0.031	&	0.861	\\
Iron	&	0.389	&	0.390	&	0.001	&	0.971	\\
\hline
	\end{longtable}
Source: Authors' estimations based on GHS Data
\end{center}

\newpage
\begin{center}
\begin{longtable}[!h]{lcccc}
\caption{Per Capita Expenditure Prediction Models, overall sample and sample with attrition (OLS)}\label{pcexpdr_attrition}
\\
 \hline
& \multicolumn{2}{c}{Sample with Attrition} &  \multicolumn{2}{c}{Overall Sample}   \\
 & (1) & (2) & (3) & (4) \\
VARIABLES & coef & se & coef & se \\ \hline
\multicolumn{5}{l}{\textbf{Social and Economic Variables}}\\
Household Size  	&	-7,941***	&	-482.8	&	-8,151***	&	-465.2	\\
Share of Dependent Members 	&	-26,090***	&	-4,074	&	-26,289***	&	-3,963	\\
State Fixed Effect 	&  \multicolumn{4}{c}{yes}\\								
Type of Settlement (Rural=1) 	&	2,011	&	-3,112	&	2,841	&	-3,000	\\
HH Gender (1=Male) 	&	23,215***	&	-6,538	&	20,081***	&	-5,794	\\
HH Age 	&	27.81	&	-66.04	&	20.13	&	-65.94	\\
HH Years of Education 	&	1,149***	&	-184.4	&	1,303***	&	-181.8	\\
HH Employed 	&	288.9	&	-4,596	&	271.3	&	-4,579	\\
HH Married 	&	6,884	&	-11,319	&	1,742	&	-10,388	\\
HH Gender * HH Married 	&	-38,980***	&	-12,082	&	-32,311***	&	-10,501	\\
HH Language  & & & & \\
\multicolumn{1}{r}{(baseline= Hausa)} & & & & \\
\multicolumn{1}{r}{Igbo}	&	-5,250	&	-7,139	&	-6,390	&	-6,910	\\
\multicolumn{1}{r}{Yoruba}	&	-3,696	&	-5,314	&	-2,704	&	-5,409	\\
\multicolumn{1}{r}{Other}	&	-1,258	&	-3,195	&	-2,030	&	-3,169	\\	
\multicolumn{5}{l}{\textbf{Quality of Dwelling}}\\						
Number of rooms 	&	1,705***	&	-504.3	&	1,753***	&	-468.4	\\
Quality of Roof   	& & & & \\							
\multicolumn{1}{r}{(baseline=Low)} 	& & & & \\							
\multicolumn{1}{r}{Medium} 	&	6,209***	&	-2,269	&	4,346**	&	-2,168	\\
\multicolumn{1}{r}{High} 	&	-2,701	&	-10,023	&	-6,430	&	-8,746	\\
Quality of Wall   		& & & & \\						
\multicolumn{1}{r}{(baseline=Low)} 	& & & & \\							
\multicolumn{1}{r}{Medium} 	&	225.2	&	-2,976	&	-507.8	&	-2,833	\\
\multicolumn{1}{r}{High}  	&	1,321	&	-2,605	&	941.9	&	-2,619	\\
Quality of Floor  		& & & & \\						
\multicolumn{1}{r}{(baseline=Low)} 		& & & & \\							
\multicolumn{1}{r}{Medium} 	&	-2,200	&	-2,597	&	-1,068	&	-2,545	\\
\multicolumn{1}{r}{High} 	&	573.7	&	-2,645	&	1,534	&	-2,737	\\
Type of Toilet & & & & \\								
\multicolumn{1}{r}{(baseline=No Facility)} 	& & & & \\								
\multicolumn{1}{r}{Flush Toilet}  	&	16,712***	&	-4,927	&	17,853***	&	-5,157	\\
\multicolumn{1}{r}{Improved Pit Latrine}  	&	-4,038	&	-5,750	&	-1,774	&	-6,563	\\
\multicolumn{1}{r}{Uncovered Pit Latrine}  	&	4,931*	&	-2,705	&	5,148*	&	-2,658	\\
\multicolumn{1}{r}{Other}  	&	5,618*	&	-3,325	&	5,083	&	-3,291	\\
Main source of lighting  	& & & & \\								
\multicolumn{1}{r}{(baseline=Electricity/Gas)} 		& & & & \\							
\multicolumn{1}{r}{Kerosene}  	&	460.7	&	-2,876	&	1,533	&	-2,835	\\
\multicolumn{1}{r}{Candles/Batteries}  	&	2,693	&	-3,471	&	2,451	&	-3,373	\\
\multicolumn{1}{r}{Other}  	&	-1,926	&	-2,882	&	-1,960	&	-2,692	\\
Main heat source for cooking 		& & & & \\							
\multicolumn{1}{r}{(baseline=Firewood)} 	& & & & \\								
\multicolumn{1}{r}{Charcoal} 	&	-3,786	&	-6,211	&	-1,354	&	-5,766	\\
\multicolumn{1}{r}{Kerosene/Oil} 	&	7,966***	&	-2,853	&	6,802**	&	-2,796	\\
\multicolumn{1}{r}{Electricity/Gas} 	&	32,053**	&	-12,680	&	32,040***	&	-12,194	\\
\multicolumn{1}{r}{Other} 	&	13,071*	&	-6,961	&	12,845**	&	-5,687	\\
Type of garbage disposal 		& & & & \\							
\multicolumn{1}{r}{(baseline=Bins)} 	& & & & \\								
\multicolumn{1}{r}{Compound}  	&	-6,878	&	-4,986	&	-10,325**	&	-5,095	\\
\multicolumn{1}{r}{Empty Lots/streets/rivers}  	&	-6,616	&	-5,013	&	-9,727*	&	-5,054	\\
\multicolumn{5}{l}{\textbf{Assets}}\\									
TV 	&	8,951***	&	-2,368	&	10,813***	&	-2,282	\\
Fridge 	&	3,858	&	-3,024	&	2,408	&	-3,106	\\
Stove 	&	6,867***	&	-2,470	&	6,178***	&	-2,366	\\
Bicycle 	&	1,321	&	-2,063	&	1,776	&	-2,046	\\
Car 	&	19,695***	&	-4,281	&	19,015***	&	-4,147	\\
Iron 	&	6,645***	&	-1,888	&	7,761***	&	-1,945	\\
	&		&		&		&		\\
Constant 	&	120,075***	&	-14,477	&	124,696***	&	-15,740	\\
	&		&		&		&		\\
Observations 	&	\multicolumn{2}{c}{4,545}		&	\multicolumn{2}{c}{4,925}		\\
 R-squared 	&	\multicolumn{2}{c}{0.453}		&	\multicolumn{2}{c}{0.462}		\\
\hline
\end{longtable}
Source: Authors' estimations based on GHS Data
\end{center}

\newpage
\begin{center}
\begin{longtable}[!h]{lcc}
\caption{Logit Model on the probability of being risk-averse (0-risk-tolerant 1-risk-averse)}\label{risk_love_logit}
\\
 \hline
 & (1) & (2) \\
VARIABLES & coef & se \\
\hline
\multicolumn{3}{l}{\textbf{Social and Economic Variables}}\\
Per Capita Expenditure 	&	 -1.79e-06** 	&	 (8.92e-07) \\
Household Size 	&	0.00192	&	 (0.0180) \\
Share of Dependent Members  	&	-0.116	&	 (0.174) \\
State Fixed Effect &	 \multicolumn{2}{c}{yes}\\			
Type of Settlement (Rural=1) 	&	0.0186	&	 (0.120) \\
HH Gender (1=Male)  	&	 -0.458** 	&	 (0.205) \\
HH Age	&	-0.00376	&	 (0.00306) \\
HH Years of Education 	&	0.000545	&	 (0.0107) \\
HH Employed 	&	 -0.558*** 	&	 (0.153) \\
HH Married 	&	-0.669	&	 (0.428) \\
HH Gender * HH Married 	&	 0.871** 	&	 (0.437) \\
HH Language			& & \\	
\multicolumn{1}{r}{(baseline= Hausa)} &&\\
\multicolumn{1}{r}{Igbo}	&	-0.33	&	 (0.285) \\
\multicolumn{1}{r}{Yoruba}	&	-0.0613	&	 (0.258) \\
\multicolumn{1}{r}{Other}	&	-0.0399	&	 (0.162) \\
\multicolumn{3}{l}{\textbf{Quality of Dwelling}}\\				
Number of Rooms  	&	-0.036	&	 (0.0232) \\
Quality of Roof  		& & \\		
\multicolumn{1}{r}{(baseline=Low)} 		& & \\		
\multicolumn{1}{r}{Medium} 	&	-0.102	&	 (0.142) \\
\multicolumn{1}{r}{High} 	&	-0.463	&	 (0.344) \\
Quality of Wall  		& & \\		
\multicolumn{1}{r}{(baseline=Low)} 		& & \\		
\multicolumn{1}{r}{Medium}  	&	 0.294* 	&	 (0.173) \\
\multicolumn{1}{r}{High}  	&	-0.00575	&	 (0.137) \\
Quality of Wall  				& & \\
\multicolumn{1}{r}{(baseline=Low)} 		& & \\		
\multicolumn{1}{r}{Medium}   	&	0.0472	&	 (0.168) \\
\multicolumn{1}{r}{High}   	&	0.098	&	 (0.165) \\
Type of Toilet 	& & \\
\multicolumn{1}{r}{(baseline=No Facility)} 	& & \\			
\multicolumn{1}{r}{Flush Toilet}  	&	-0.147	&	 (0.207) \\
\multicolumn{1}{r}{Improved Pit Latrine}  	&	0.0731	&	 (0.244) \\
\multicolumn{1}{r}{Uncovered Pit Latrine}  	&	-0.19	&	 (0.139) \\
\multicolumn{1}{r}{Other}  	&	-0.133	&	 (0.174) \\
Main source of lighting  		& & \\		
\multicolumn{1}{r}{(baseline=Electricity/Gas)} 	& & \\			
\multicolumn{1}{r}{Kerosene} 	&	-0.0824	&	 (0.118) \\
\multicolumn{1}{r}{Candles/Batteries} 	&	0.0745	&	 (0.173) \\
\multicolumn{1}{r}{Other} 	&	 -0.386** 	&	 (0.164) \\
Main heat source for cooking 		& & \\		
\multicolumn{1}{r}{(baseline=Firewood)} 	& & \\			
\multicolumn{1}{r}{Charcoal} 	&	 1.286** 	&	 (0.592) \\
\multicolumn{1}{r}{Kerosene/Oil} 	&	0.12	&	 (0.142) \\
\multicolumn{1}{r}{Electricity/Gas} 	&	-0.111	&	 (0.397) \\
\multicolumn{1}{r}{Other} 	&	 0.392* 	&	 (0.232) \\
Type of garbage disposal 			& & \\	
\multicolumn{1}{r}{(baseline=Bins)} 	& & \\			
\multicolumn{1}{r}{Compound} 	&	0.12	&	 (0.175) \\
\multicolumn{1}{r}{Empty Lots/streets/rivers} 	&	0.0721	&	 (0.165) \\
\multicolumn{3}{l}{\textbf{Assets}}\\				
TV 	&	-0.129	&	 (0.122) \\
Fridge 	&	0.0795	&	 (0.140) \\
Stove 	&	-0.135	&	 (0.119) \\
Bicycle 	&	0.143	&	 (0.104) \\
Car 	&	0.0267	&	 (0.162) \\
Iron 	&	-0.0684	&	 (0.116) \\
Constant 	&	 1.436*** 	&	 (0.522) \\
 	&	  	&	  \\
 Observations 	&	4,545	&	  \\ \hline
\multicolumn{3}{c}{ Robust standard errors in parentheses} \\
\multicolumn{3}{c}{ *** p$<$0.01, ** p$<$0.05, * p$<$0.1} \\
\end{longtable}
Source: Authors' estimations based on GHS Data
\end{center}

\newpage
\begin{center}
\begin{longtable}[!h]{lcc}
\caption{Expenditure Prediction Model (Logaritm of total per-capita household expenditure) for households living in non-conflict areas (OLS)}\label{prediction}
\\
 \hline
 & (1) & (2) \\
VARIABLES & coef & se \\
\hline
\multicolumn{3}{l}{\textbf{Social and Economic Variables}}\\
Household Size 	&	-0.0949***	&	(-0.00309)	\\
Share of Dependent Members 	&	-0.286***	&	(-0.0247)	\\
State Fixed Effect 	& 	 \multicolumn{2}{c}{yes}\\			
Type of Settlement (Rural=1) 	&	0.0332	&	(-0.0233)	\\
HH Gender (1=Male) 	&	0.201***	&	(-0.0289)	\\
HH Age 	&	0.00313***	&	(-0.000531)	\\
HH Years of Education 	&	0.0236***	&	(-0.00162)	\\
HH Employed 	&	0.128***	&	(-0.0179)	\\
HH Married 	&	-0.0806	&	(-0.0529)	\\
HH Gender * HH Married 	&	-0.189***	&	(-0.0601	)\\
HH Language		&&\\			
 \multicolumn{1}{r}{(baseline= Hausa)} &&\\					
\multicolumn{1}{r}{Igbo}	&	0.0117	&	(-0.0516)	\\
\multicolumn{1}{r}{Yoruba}	&	-0.0349	&	(-0.0473)	\\
\multicolumn{1}{r}{Other}	&	-0.0325	&	(-0.0358)	\\
\multicolumn{3}{l}{\textbf{Quality of Dwelling}}\\					
Number of Rooms 	&	0.0346***	&	(-0.00347)	\\
Quality of Roof  		&&\\			
\multicolumn{1}{r}{(baseline=Low)} 		&&\\			
\multicolumn{1}{r}{Medium}		&	0.0656***	&	(-0.0191)	\\
\multicolumn{1}{r}{High}			 &	0.0231	&	(-0.0536)	\\
Quality of Wall  			&&\\		
\multicolumn{1}{r}{(baseline=Low)} 		&&\\			
\multicolumn{1}{r}{Medium} 	&	0.00103	&	(-0.0256)	\\
\multicolumn{1}{r}{High} 	&	-0.0255	&	(-0.0163)	\\
Quality of Floor   		&&\\			
\multicolumn{1}{r}{(baseline=Low)} 	&&\\				
\multicolumn{1}{r}{Medium} 	&	0.0343	&	(-0.0257)	\\
\multicolumn{1}{r}{High} 	&	0.125***	&	(-0.027)	\\
Type of Toilet 					&&\\
\multicolumn{1}{r}{(baseline=No Facility)} 	&&\\				
\multicolumn{1}{r}{Flush Toilet} 	&	0.234***	&	(-0.0266)	\\
\multicolumn{1}{r}{Improved Pit latrine} 	&	-0.0628	&	(-0.0563)	\\
\multicolumn{1}{r}{Uncovered Pit Latrine} 	&	0.0431**	&	(-0.0209)	\\
\multicolumn{1}{r}{Other} 	&	0.0262	&	(-0.03)	\\
Main source of lighting  		&&\\			
\multicolumn{1}{r}{(baseline=Electricity/Gas)} &&\\					
\multicolumn{1}{r}{Kerosene} 	&	-0.0832***	&	(-0.016)	\\
\multicolumn{1}{r}{Candles/Battery} 	&	-0.0583***	&	(-0.0222)	\\
\multicolumn{1}{r}{Other} 	&	-0.122***	&	(-0.0234)	\\
Main heat source for cooking 		&&\\			
\multicolumn{1}{r}{(baseline=Firewood)} &&\\					
\multicolumn{1}{r}{Charcoal} 	&	0.0625	&	(-0.0486)	\\
\multicolumn{1}{r}{Kerosene/Oil} 	&	0.130***	&	(-0.0197)	\\
\multicolumn{1}{r}{Electricity/Gas} 	&	0.283***	&	(-0.0392)	\\
\multicolumn{1}{r}{Other} 	&	0.339***	&	(-0.0501)	\\
Type of garbage disposal 		&&\\			
\multicolumn{1}{r}{(baseline=Bins)} &&\\					
\multicolumn{1}{r}{Compound} 	&	-0.0722***	&	(-0.0233)	\\
\multicolumn{1}{r}{Empty Lots/streets/rivers} 	&	-0.0560***	&	(-0.0191)	\\
\multicolumn{3}{l}{\textbf{Assets}}\\					
TV 	&	0.0408***	&	(-0.0155)	\\
Fridge 	&	0.0259	&	(-0.0167)	\\
Stove 	&	-0.00359	&	(-0.0161)	\\
Bicycle 	&	-0.0800***	&	(-0.015)	\\
Car 	&	0.0593***	&	(-0.0208)	\\
Iron 	&	0.00704	&	(-0.0134)	\\
					
Constant 	&	11.33***	&	(-0.0847)	\\
 	&	&	\\		
Observations 	&	21,271	&	\\	
 R-squared 	&	0.473	&	\\ \hline	
\multicolumn{3}{c}{ Robust standard errors in parentheses} \\
\multicolumn{3}{c}{ *** p$<$0.01, ** p$<$0.05, * p$<$0.1} \\
\end{longtable}
Source: Authors' estimations based on GHS and ACLED Data
\end{center}

\newpage
\begin{table}
\caption{Association between migration and risk-aversion (Probit)}\label{prediction_risk}
\begin{tabular}{lccc} \hline\label{risk}
 & (1) & (2) & (3) \\
VARIABLES & \multicolumn{1}{l}{0 = HH migr.} & \multicolumn{1}{l}{0 = HH migr.} & \multicolumn{1}{l}{0 = HH migr.} \\
			 & \multicolumn{1}{l}{1 = HH not migr.} & \multicolumn{1}{l}{1 = HH not migr.}  & \multicolumn{1}{l}{1 =HH not migr.}  \\
 & 				 & \multicolumn{1}{l}{and some conflict}  & \multicolumn{1}{l}{and always conflict} \\
 \hline
 &  &  &  \\
 \multicolumn{4}{l}{\textbf{Behavioural}}\\
Risk Aversion 	&	 -0.179** 	&	 -0.363*** 	&	 -0.400** \\
 \multicolumn{4}{l}{\textbf{Social and Economic Variables}}\\						
Household Size 	&	 0.0324** 	&	 0.0465** 	&	 0.153** \\
Share of Dependent 	&	 -0.398*** 	&	 -0.593*** 	&	 -0.00753 \\
HH Gender = Male 	&	0.104	&	0.201	&	 0.209 \\
HH Age 	&	 0.0158*** 	&	 0.0188*** 	&	 0.0165*** \\
HH Years of Education 	&	 -0.0273*** 	&	 -0.0369*** 	&	 -0.0426* \\
HH Employed 	&	 0.364*** 	&	 0.503*** 	&	 0.877*** \\
HH Married 	&	0.188	&	0.0615	&	 1.192*** \\
HH Gender * HH Married 	&	-0.21	&	-0.221	&	 -1.532** \\
 \multicolumn{4}{l}{\textbf{Quality of Dwelling}}\\						
Number of Rooms 	&	 0.0614* 	&	0.0527	&	 -0.0260 \\
 \multicolumn{4}{l}{Quality of Walls}\\						
\multicolumn{1}{r}{(baseline=Low)} 	&		&	&	\\	
\multicolumn{1}{r}{Medium} 	&	-0.175	&	-0.314	&	 0.509* \\
\multicolumn{1}{r}{High} 	&	 -0.444*** 	&	 -0.564*** 	&	 -0.254 \\
 \multicolumn{4}{l}{\textbf{Assets}}\\						
TV 	&	-0.0738	&	-0.0483	&	 0.382 \\
Fridge 	&	0.078	&	 0.259* 	&	 -0.527 \\
Stove 	&	 -0.273*** 	&	 -0.372*** 	&	 -0.939** \\
Bicycle 	&	 0.354*** 	&	 0.408* 	&	 0.368 \\
Car 	&	 -0.280** 	&	 -0.486*** 	&	 -0.319 \\
Iron 	&	0.0453	&	0.113	&	 -0.260 \\
	&		&		&	\\
Constant 	&	 0.819*** 	&	0.0513	&	 -2.585*** \\
 	&	  	&	  	&	  \\
Observations 	&	4,577	&	1,289	&	 359 \\
 Pseudo R-squared 	&	0.141	&	0.19	&	 0.357 \\ \hline
\multicolumn{3}{c}{ Robust standard errors (not reported)} \\
\multicolumn{4}{c}{ *** p$<$0.01, ** p$<$0.05, * p$<$0.1} \\
\end{tabular}
Source: Authors' estimations based on GHS and ACLED Data\\
\end{table}

\begin{table}
\caption{Difference in Differences Test (OLS)}\label{prediction_did}
\begin{center}
\begin{tabular}{lcc} \hline
 & (1) & (2) \\
	&	Risk Aversion	&	Risk Aversion	\\
	&	&	\\		
Migration Group (1=never migrated) 	&	-0.105***	&	-0.0873**	\\
Conflict (1=some conflict) 	&	-0.049	&	-0.0501	\\
did 	&	-0.0124	&	-0.00969	\\
 \multicolumn{3}{l}{\textbf{Social and Economic Variables}}\\					
Household Size 	&	&	0.000533	\\	
Share of Dependent 	&	&	0.0426	\\	
HH Gender = Male 	&	&	-0.0172	\\	
HH Age 	&	&	-0.000869	\\	
HH Years of Education 	&	&	-0.00298	\\	
HH Employed 	&	&	-0.023	\\	
HH Married 	&	&	-0.055	\\	
HH Gender * HH Married 	&	&	0.0738	\\	
 \multicolumn{3}{l}{\textbf{Quality of Dwelling}}\\					
Number of Rooms 	&	&	-0.0146***	\\	
 \multicolumn{3}{l}{Quality of Walls}\\					
\multicolumn{1}{r}{(baseline=Low)} & &	\\				
\multicolumn{1}{r}{Medium} 	&	&	0.0649*	\\	
\multicolumn{1}{r}{High} 	&	&	0.0128	\\	
 \multicolumn{3}{l}{\textbf{Assets}}\\					
TV 	&	&	0.0187	\\	
Fridge 	&	&	0.0147	\\	
Stove 	&	&	-0.0222	\\	
Bicycle 	&	&	-0.0406*	\\	
Car 	&	&	0.00511	\\	
Iron 	&	&	0.0198	\\	
Constant	&	0.820***	&	0.905***	\\
	&	&	\\		
Observations	&	4,577	&	4,577	\\
R-squared	&	0.006	&	0.016	\\ \hline
\multicolumn{3}{c}{ Robust standard errors (not reported)} \\
\multicolumn{3}{c}{ *** p$<$0.01, ** p$<$0.05, * p$<$0.1} \\
\end{tabular}
\end{center}
Source: Authors' estimations based on GHS and ACLED Data
\end{table}
\end{document}